\documentclass[12pt]{article}

\usepackage{amsfonts}

\usepackage{amssymb}
\usepackage{latexsym}
\usepackage{graphicx}
\usepackage[english]{babel}

\usepackage{amsfonts}
\usepackage{latexsym}
\usepackage{graphicx}
\usepackage[english]{babel}
\begin{document}
\input epsf

\def\p{\partial}
\def\h{{1\over 2}}
\def\be{\begin{equation}}
\def\bea{\begin{eqnarray}}
\def\ee{\end{equation}}
\def\eea{\end{eqnarray}}
\def\d{\partial}
\def\la{\lambda}
\def\eps{\epsilon}
\def\bb{\bigskip}
\def\mm{\medskip}
\newcommand{\dm}{\begin{displaymath}}
\newcommand{\edm}{\end{displaymath}}
\renewcommand{\b}{\tilde{B}}
\newcommand{\gm}{\Gamma}
\newcommand{\ac}[2]{\ensuremath{\{ #1, #2 \}}}
\renewcommand{\ell}{l}
\newcommand{\z}{\ell}
\newcommand{\newsection}[1]{\section{#1} \setcounter{equation}{0}}
\def\bb{$\bullet$}
\def\Qbar{{\bar Q}_1}
\def\QPbar{{\bar Q}_p}

\def\q{\quad}

\def\bn{B_\circ}

\let\a=\alpha \let\b=\beta \let\g=\gamma \let\d=\delta \let\e=\epsilon
\let\c=\chi \let\th=\theta  \let\k=\kappa
\let\l=\lambda \let\m=\mu \let\n=\nu \let\x=\xi \let\r=\rho
\let\s=\sigma \let\t=\tau
\let\vp=\varphi \let\vep=\varepsilon
\let\w=\omega      \let\G=\Gamma \let\D=\Delta \let\Th=\Theta
                     \let\P=\Pi \let\S=\Sigma

\def\h{{1\over 2}}
\def\t{\tilde}
\def\r{\rightarrow}
\def\nn{\nonumber\\}
\let\bm=\bibitem
\def\Kt{{\tilde K}}
\def\b{\bigskip}

\let\p=\partial

\begin{flushright}
\end{flushright}
\vspace{20mm}
\begin{center}
{\LARGE  Effective information loss outside the horizon\footnote{(Slightly expanded version of ) essay written for the Gravity Research Foundation 2011 essay contest.}}
\\
\vspace{18mm}
{\bf  Samir D. Mathur }\\

\vspace{8mm}
Department of Physics,\\ The Ohio State University,\\ Columbus,
OH 43210, USA\\mathur@mps.ohio-state.edu
\vspace{4mm}
\end{center}
\vspace{10mm}
\thispagestyle{empty}
\begin{abstract}

If a system falls through a black hole horizon, then its information is lost to an observer at infinity. But we argue that the  {\it accessible} information  is lost {\it before} the horizon is crossed. The temperature of the hole limits information carrying signals from a system that has fallen too close to the horizon. Extremal holes have T=0, but there is a minimum energy required to emit a quantum in  the short proper time left before the horizon is crossed. If we attempt to bring the system back to infinity for observation, then acceleration radiation destroys the information. All three considerations give a critical distance from the horizon
$d\sim \sqrt{r_H\over \Delta E}$, where $r_H$ is the horizon radius and $\Delta E$ is the energy scale characterizing the system.   For systems in string theory where we pack information as densely as possible,  this acceleration constraint is found to have a geometric interpretation. These estimates suggest that in theories of  gravity we should measure information  not as a quantity contained inside a given system, but in terms of how much of that information  can be reliably accessed by another observer.

\end{abstract}
\vskip 1.0 true in

\newpage
\setcounter{page}{1}

\section{Introduction}

The notions of gravity, entropy and information form a closely linked triangle. It is plausible that unlocking the mysterious relations in  this triangle will provide a deep insight into the nature of spacetime and  quantum theory. 

Consider a system that falls towards a black hole. If the system crosses  the horizon, then its information will be lost to an outside observer. But in some approaches to the black hole problem, the region outside the hole is supposed to give a complete and self-consistent description of physics. The idea of black hole complementarity is one such situation, where one assumes that the infalling observer would get destroyed at the horizon (and have its information re-radiated to infinity); it is only in a second complementary description that he falls through the horizon \cite{thooft1,suss1}. Recently  efforts have been directed to showing that the backreaction of Hawking radiation may be severe enough to prevent a shell from falling through its horizon \cite{vachas}. In string theory one finds that black hole microstates do not have regular horizons; instead they are `fuzzballs' \cite{fuzz}. 

In such situations, one may think that the information in an infalling bit would be preserved until the infalling object reaches the horizon (or the surface of the fuzzball). But in this essay we will show that the information available to the observer at infinity actually starts losing its fidelity as the system falls towards the horizon, and is effectively wiped out before the horizon is crossed. 

Consider a system having mass $m$, falling radially  from rest at infinity, towards a black hole with metric
\be
ds^2=(1-{2M\over r})dt^2+{1\over 1-{2M\over r}}\,  dr^2+r^2\, d\Omega_2^2
\label{one}
\ee
When the system has reached a radius $\bar r$, its inward proper velocity is $U^r={dr\over d\tau}=-\sqrt{2M\over \bar r}$. For $\bar r \approx 2M$, we get $U^r\approx -1$, so the system is falling in very fast indeed. Now suppose  the system tries to send the information about its state to the observer at infinity, by encoding this information in the spin of an emitted photon.  By local momentum conservation, at most $\h m$ of the system's energy can be given to this photon. This photon is redshifted as it reaches infinity, arriving with an energy 
\be
E_\gamma\approx {m\over 4}\, {(\bar r-2M)\over 2M}
\ee
But if $E_\gamma\lesssim kT$, where $T$ is the temperature of the hole, then this information carrying photon cannot be distinguished from the photons in the thermal bath of Hawking radiation, and we cannot hope to recover the information of the system. Thus we find that the information in the system is effectively lost when
\be
{\bar r -2M\over 2M}\sim {4kT\over m}
\label{two}
\ee
As an example, if the system has energy $m=20\, kT$, then this condition is  $\bar r-2M \sim 0.4M$.

But one may argue that we do not need to have the system send back its information using its own internal energy; the observer at infinity is in possession of infinite resources, and can just send a device to scoop up the system and bring it back to infinity, where he can measure its state at leisure. Here we will encounter the following difficulty:  to prevent the system at position $\bar r$ (and $U^r\approx -1$) from falling past the horizon, we need some minimum acceleration $a$ to be maintained for a certain proper time $\Delta \tau$. This acceleration leads to Unruh radiation being felt by the system \cite{unruh}, and its information carrying bits can then get knocked out of their state into new states. Thus we again find a degradation of the information in the system. 

\begin{figure}[htbp]
\begin{center}
\includegraphics[scale=.38]{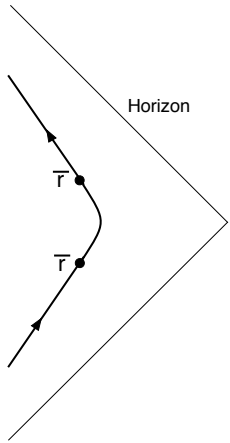}
\caption{{The trajectory of the system in the $t-r$ plane. We consider free infall until the position $\bar r$, then a period $\Delta\tau$ of constant acceleration $a$ which returns us to $\bar r$ with velocity reversed,  then free motion again out to infinity.}}
\label{fetwo}
\end{center}
\end{figure}

To compute this effect, we let the information be encoded in a 2-level system with energy gap $\Delta E$ and a coupling $H_{int}=g\hat O\hat\phi$ to a massless scalar field $\phi$. Letting $g\langle\psi_2 | \hat O |\psi_1\rangle\equiv \alpha$, we find that the probability per unit proper time for the system to get exited by absorbing a $\hat \phi$ quantum from the Rindler bath is
\be
\Gamma={|\alpha|^2 \Delta E\over 2\pi}{1\over e^{2\pi a^{-1}\Delta E}-1}
\ee
We let the system be in free fall till radius $\bar r$ as before, then  maintain a constant $a$ until the time that we return at position $\bar r$ with the sign of $U^r$ reversed; then we  let the system coast back to infinity under free motion. For such a path (shown in fig.\ref{fetwo}) we find that \cite{mathurnew}
\be
\Delta \tau={2\over a}\, \sinh^{-1}{1\over 2[(\bar r-2M)-a^{-1}]}
\ee
The probability of excitation is $P=\Gamma\Delta \tau$, and we should choose $a$ to minimize $P$.  If we take $a$ too large (sudden turnaround) then we are almost certain to change the state of the 2-level system. If we take $a\lesssim [4(\bar r-2M)]^{-1}$, then we fall through the horizon. Assuming $|\alpha|^2$ is not parametrically different from unity,  one finds that $P\sim 1$ if
\be
\bar r-2M \sim {1\over \Delta E}
\label{el}
\ee
Let us compare this to (\ref{two}). If we let the energy scale of the system be $\Delta E\sim m$, and note that $T={1\over 8\pi M}$, then we find that the two estimates of the critical radius are of the same order. 

One may wonder if there is an estimate of type (\ref{two}) for extremal holes
\be
ds^2=-(1-{M\over r})^2 dt^2+{dr^2\over (1-{M\over r})^2}+r^2 d\Omega_2^2
\ee
which have temperature $T=0$. 
 Again consider infall of a system of mass $m$, starting from rest at infinity, and reaching a point $\bar r$ with ${\bar r-M\over M}\ll 1$. We have $U^r={dr\over d\tau}\approx -1$ as before, and this allows only a proper time
\be
\Delta\tau\sim {\bar r-M}
\ee
for the system to emit a photon to infinity before it falls through the horizon. But a quantum emitted in this short time must have an  energy
$\Delta E\gtrsim {1\over \Delta\tau}$, and setting $\Delta E\sim m$ we find that the closest we come to the horizon before losing information is
\be
\bar r-M\sim {1\over m}
\label{twop}
 \ee
For the neutral hole, this argument gives $\bar r-2M\sim {1\over m}$, which agrees with (\ref{two}).

 It is interesting that all three ways of estimating the critical distance give the same result. 
We can convert the coordinate distance   $\bar r-2M$  for the neutral hole to a proper distance $d$ using the metric (\ref{one}), finding for this critical distance
\be
d\sim \sqrt{r_H\over \Delta E}
\ee
where $r_H=2M$ is the Schwarzschild radius and $\Delta E$ is the energy scale characterizing the system. For the extremal hole the proper distance to the horizon is in fact infinite, which suggests a general inequality $d\gtrsim \sqrt{r_H\over \Delta E}$.

It remains true however that the critical radius we have obtained depends on the details $\Delta E, \alpha$  characterizing the system. If we are to hope for a universal condition, then we should make our system using the objects present in a full theory of quantum gravity; in this case the couplings and energy levels would be presumably related to the gravitational coupling and density of states in gravity. 

Thus consider string theory, where we compactify 6 directions on small circles. For our infalling system we take a string wrapped $n_1$ times around one of the circles. We add $n_p$ units of momentum along the string; this momentum is carried by transverse vibrations $w(t-y)$ where $y$ is the coordinate along the circle. Since the momentum can be partitioned among different harmonics in many ways, we find   $N=Exp[S_{ex}]$ possible states, where
\be
S_{ex}=2\sqrt{2}\pi\sqrt{n_1n_p}
\ee
turns out to agree with the expression for the entropy of an extremal black hole with charges $n_1, n_p$ \cite{sen}. Thus this vibrating string - called the NS1-P system in string theory -- allows us to store information very densely;  choosing one of the allowed $N$ states stores $\ln N=S_{ex}$ bits of information. It is plausible that systems like these (whose entropy agrees with the entropy of a black hole with the same mass and charges) would be the most dense information storing devices in any complete theory of quantum gravity. 

This time we know the energy levels of our system -- they are just the vibration levels of the string, described by left and right moving transverse deformations $w(t\pm y)$.  For the scalar field $\phi$ we can take any one of the gravitons $h_{ij}$ with indices in the 6 compact directions. The coupling of these scalars to the vibrations has the form $S_{int}=C\int dt dy ~\phi \, \p_{t-y}w\, \p_{t+y}w$, where the constant $C$ was computed in \cite{dasmathur}.
When our system accelerates, the coupling $S_{int}$ can excite it to any other energy level, so we have to perform a weighted sum over final energy levels, getting a factor $ \int dE_f \, \rho(E_f)\, |\alpha_{i\r f}|^2\equiv Q$. 

We now have all the tools available to compute the excitation rate, but at this point we note that the result of this computation can be already extracted from known results. Suppose we send a beam of $\phi$ quanta onto our NS1-P system in flat space, and compute the absorption cross section $\sigma$. The couplings and level density again appear in the combination $Q$. Thus for our accelerating system we can package all the needed system characteristics into its cross section $\sigma$. But $\sigma$ for the NS1-P system was computed in \cite{dasmathur}, and has a surprisingly simple form; we have $\sigma=A_{ex}$, where $A_{ex}$ is the horizon area of a black hole with charges $n_1, n_p$. Putting these results into the above computation of excitation by acceleration, the critical radius for information degradation ($P\sim 1$) is found to be $\bar r-2M\sim r_{ex}$, where  $r_{ex}\sim A_{ex}^\h$ is the radius of the small extremal hole that would have charges $n_1, n_p$ in string theory.  Converting the  coordinate range $\bar r-2M$ to a proper distance $d$ from the horizon using (\ref{one}) we get
\be
d\sim \sqrt{r_{ex}r_H}\gg r_{ex}
\label{ten}
\ee
where $r_H=2M$ is the Schwarzschild radius of the hole, and  the inequality holds when we take $r_{ex}\ll r_H$. 
The result (\ref{ten}) is found to hold even if we use many other kinds of charges and consider holes in different dimensions $D$. 

Thus we see that if we take a system that   packs information as densely as possible in a full theory of gravity, then the critical radius determined by acceleration radiation  is much larger than the distance of approach where the system would physically touch the horizon; thus 
we do lose information `outside' the horizon.

These results  suggest a change in our perspective on information in theories of gravity. First, we should think in terms of  how much information we can store for {\it given energy}. Second,  we should measure information not as a quantity contained in a given system, but in terms of how much of that information  can be reliably {\it accessed} by another observer. If the system and the observer are separated by a gravitational potential, then there is a reduction in the fidelity of the information of the system from the viewpoint of the observer, even though the system has not crossed a horizon. This degradation of information becomes infinitely strong as a horizon is approached, so we should not think of the information as lost `suddenly' when the horizon is crossed, but as a gradual loss when the horizon is approached. In fact we do not need a horizon to have  information degradation, which ties up well with the discovery in string theory that energy eigenstates do not form horizons, but rather form horizon sized quantum `fuzzballs'.

\section*{Acknowledgements}

This work was supported in part by DOE grant DE-FG02-91ER-40690.

\end{document}